\documentclass[%
 reprint,
 amsmath,amssymb,
 aps,
]{revtex4-1}

\usepackage{graphicx}% Include figure files
\usepackage{dcolumn}% Align table columns on decimal point
\usepackage{bm}% bold math
\begin{document}

\preprint{APS/123-QED}

\title{How does active participation effect consensus: Adaptive network model of opinion dynamics and influence maximizing rewiring}% Force line breaks with \\

\author{Markus Brede}
\affiliation{%
 Department of Electronics and Computer Science\\
 University of Southampton, UK %\textbackslash\textbackslash
}%


%\author{Markus Brede}
%\affiliation{
%\affiliation{
 %Third institution, the second for Charlie Author
%\affiliation{%


\date{\today}% It is always \today, today,

\begin{abstract}
In this paper we study the impact of active participation -- or deliberately seeking out other agents with an aim to convince them -- on the dynamics of consensus formation. For this purpose, we propose an adaptive network model in which two processes shape opinion dynamics at interwoven time-scales: (i) agents adapt their opinions subject to influence from social network neighbours who hold opinions within a tolerance interval $\delta$ and (ii) agents rewire network connections with an aim to maximize their own influence on overall system opinion. We study this system in both an endogenous setting in which all agents are subject to influence and also attempt to maximize influence, and in a setting of exogenous control, in which external agents not subject to influence adaptively attempt to maximize their influence. In both settings we find three regimes of stationary opinion configurations: (i) for low $\delta$ a regime of two evenly balanced radicalized opinion clusters at the extremes of the opinion space, (ii) for intermediate $\delta$ a 'winner-takes-most' regime of two unevenly sized radicalized opinion clusters, and (iii) for large $\delta$ a regime in which very low spread compromise consensus states can be reached. Comparing to adaptive processes of random and deliberately spread-reducing rewiring, we demonstrate that in regime (iii) competitive influence maximization can achieve near-minimal opinion spread within near-optimal times. Further, we also show that competitive influence maximizing rewiring can reduce the impact of small influential minorities on consensus states.

%\begin{description}
%\end{description}
\end{abstract}

\pacs{Valid PACS appear here}% PACS, the Physics and Astronomy
\maketitle

%\tableofcontents

\section{\label{sec:level1} Introduction}

Insights about the dynamics of opinion formation are relevant for our understanding of a number of social and economic processes, ranging from studies of radicalization \cite{Ramos:2015,Galam2016:2016}, political campaigns \cite{Javarone:2014,Hegselmann:2015}, the spread of technology \cite{Laciana:2011} or the development of industries \cite{Alshamsi:2018} to applications in financial markets \cite{Kleinbergbook} and they might ultimately shed light on how democracies arrive at political decisions. The field has seen a rich interdisciplinary literature in econo-physics and computer science, cf. \cite{Castellano:2009,Sirbu:2016} for reviews. A slightly different take on models of opinion formation is to study the role of agents who actively attempt to influence the population, e.g., with the intent to influence decisions to align with particular goals. The latter setting clearly has relevant applications in advertising and viral marketing \cite{Mehmood:2016}, but might also shed light on questions about the spread of fake news or foreign influence in political elections. 

Mostly focusing on issues of optimal allocations of influence to nodes on a social network influence maximization has found much attention, with most work concentrating on variants of the independent cascade or threshold type models \cite{Kleinberg:2003} and elegant computational optimization techniques \cite{Morone:2015}, including considerations of competitive settings \cite{Bharathi:2007,Borodin:2010,Goyal:2014}.

%However, in variants of the very influential independent cascade model agents are typically committed to an opinion once an opinion has been adopted and such models thus fail to capture the rich dynamics of flipping back and forth in settings in which there is little cost to opinion change. Better suited for this scenario are dynamic models that describe the evolution of opinion updates based on stochastic rules that allow agents to change their opinions back and forth.  At a basic level one can distinguish two types of modeling approaches, models like the voter dynamics \cite{Clifford:1973,Holley:1975} and extensions to the multi-state voter model \cite{Starnini:2012} or Ising-like dynamics \cite{Galam:1982} and the related Sznajd model \cite{Sznajd:2000} that treat opinions as discrete and models like the DeGroot model \cite{DeGroot:1974} or, including considerations of bounded confidence, the models of Weissbuch and Deffuant \cite{Weissbuch:2000} and similar kinetic exchange models \cite{Lallouache:2010,Sen:2011} that treat opinions as continuous, typically studying one-dimensional opinion spaces.

%%rewrite here:
However, in variants of the very influential independent cascade model agents are typically committed to an opinion once an opinion has been adopted and such models thus fail to capture the rich dynamics of flipping back and forth in settings in which there is little cost to opinion change. Better suited for this scenario are dynamic models that describe the evolution of opinion updates based on stochastic rules that allow agents to change their opinions back and forth. At a basic level one can distinguish two types of modeling approaches: models that treat opinions as discrete and models that deal with a continuous, very often one-dimensional, opinion space. Prominent approaches in the first category are the voter dynamics \cite{Clifford:1973,Holley:1975} and extensions to the multi-state voter model \cite{Starnini:2012} or Ising-like dynamics \cite{Galam:1982} and the related Sznajd model \cite{Sznajd:2000}. In a similar vein, also related models of cultural dynamics like the Axelrod model \cite{Axelrod:1997} or studies of media influence based on it \cite{Avella:2007} have considered discrete opinion spaces. Related to our study is also the work of \cite{Laguna:2005} in which the role of hierarchical organizations and authority on consensus dynamics has been studied. Models of continuous opinion spaces are mostly derived from the DeGroot model \cite{DeGroot:1974} or, including considerations of bounded confidence, from the models of Weissbuch and Deffuant \cite{Weissbuch:2000,Fortunato:2004}  and similar kinetic exchange models \cite{Lallouache:2010,Sen:2011}. Variants of the model of Weissbuch and Deffuant have proved very influential in the literature and have recently also been used to study algorithmic biases in the context of opinion formation \cite{Sirbu:2019} or opinion dymamics on interacting networks of competing media and social networks \cite{Quatt:2014}.

External influence has mostly been modelled by introducing so-called `zealots' \cite{Mobilia:2003} that represent agents with a bias \cite{Mobilia:2003,Mobilia:2005} or even unidirectional influence \cite{Mobilia:2007} that can lead to consensus states in which multiple opinions can coexist. Zealotry in various forms has been extensively studied with some recent extensions to the noisy voter model \cite{Khalil:2018} or some models that also include the effects of mass media in models derived from the multi-state voter model \cite{Hu:2017,Hu:2017b}. However, studies of zealotry typically do not focus on the questions of optimal allocation or competition in the placement of zealots in the social system.

Recent literature has started to address questions of optimal allocation for the dynamic models of opinion formation, with studies focusing on the voter dynamics \cite{Kuhlman:2013,Yildiz:2013,Masuda:2015,Brede:2018,Brede:2019}, Ising-like models \cite{Lynn:2016}, and variants of AB models \cite{Arendt:2015}. While some studies point to a nuanced picture of optimal allocations depending on noise and details of the goal-functions of the optimizing parties \cite{Brede:2018,Brede:2018b}, a common thread is that optimal allocations are often well approximated by targeting hub nodes \cite{Kuhlman:2013,Masuda:2015}.

However, whilst optimal allocation for dynamic models has found some recent attention, up to our best knowledge adversarial settings in which competing agents strive for maximal influence for dynamic models have not yet been studied. Here, we investigate the impact of competitive influence maximization on opinion dynamics in a continuous model which we introduce as a variant of the Weissbuch-Deffuant model \cite{Weissbuch:2000} and related Hegselmann and Krause model \cite{Hegselmann:2002}. Different from most studies of influence maximization which focus on details of allocation rules for static networks, we formulate our models as an adaptive network model \cite{Blasius:2008}, in which opinion change and competitive influence-maximizing rewiring interact at similar time scales. We thus essentially explore the effects of agents' active participation in opinion dynamics via deliberately seeking out suitable other agents with an aim to convince them of their own opinions. The literature provides several models of adaptive opinion dynamics and network change for the case of network rewiring inspired by agents seeking homophily \cite{Holme:2006,Kozma:2008,Kimura:2008} which typically results in a phase transition between a multi-cluster state and a consensus state. However, up to our best knowledge ideas of competitive influence maximizing rewiring have not been studied in the context of adaptive network models so far.

Below, instead of focusing on details of optimal allocation which is the main aim of studies on influence maximization \cite{Kleinberg:2003,Morone:2015,Yildiz:2013,Kuhlman:2013,Masuda:2015,Brede:2018,Brede:2018b}, our main interest is in the impact of influence-maximizing rewiring on the overall opinion dynamics. We thus investigate the question how active participation of agents in opinion competition impacts on consensus states and equilibration time scales. As we will see below, depending on parameters of the consensus dynamics, active influence-maximizing rewiring can either lead to radicalization and a split of societies into extreme opinion clusters at the boundaries of the opinion space or be in a regime in which more aligned compromises can be reached in near-optimal time. 

Often in opinion dynamics in populations with peer influence small but well-connected minorities can have a dominating influence on consensus states \cite{Couzin:2011}. Interestingly, we also find that competitive influence maximization strongly reduces the impact of such minorities.

The remainder of the paper is organized as follows. In Sec. \ref{sec:level2} we outline our modelling choices and define metrics that quantify the opinion dynamics. Section \ref{results} then presents our main findings and with a summary and discussion in section \ref{summary} we conclude the paper.

%\cite{Kleinberg:2003,Tang:2015}, \cite{Blasius:2008}
\section{\label{sec:level2} Model}
We consider a population of $N$ agents, each of which holds an opinion $x \in[-1,1]$ described by a continuous variable with initial opinions being drawn uniformly at random from $[-1,1]$. Agents are connected by a binary directed social network given by its adjacency matrix $a_{ij}$, with $a_{ij}=1$ if $i$ links to $j$ and $a_{ij}=0$ otherwise. A connection from agent $j$ to agent $i$ models that $j$ can influence $i$. Through peer influence and external factors opinions of agents are subject to change via
\begin{equation}
\label{E0}
    x_i(t+1)= x_i(t) + \alpha \sum_j a_{ij} H_\delta (x_j(t)-x_i(t)) + \Delta \eta_i,
\end{equation}
where $H_\delta (x-y)=x-y$ if $\mid x-y\mid \leq \delta$ and $H_\delta (x-y)=0$ otherwise. In Eq. (\ref{E0}) above, $\alpha \in (0,1)$ models the rate of opinion change, $\delta$ gives a tolerance parameter, such that only agents whose opinions differ by less than $\delta$ will be taken into account when updating, and $\Delta$ gives the strength of external effects modelled by uncorrelated white noise $\eta$. According to (\ref{E0}) agents will adjust their opinions depending on the average effect of neighbours with opinions not farther away than $\delta$ and will also be affected by external factors whose weight we model by the strength of noise $\Delta$. By introducing a tolerance parameter $\delta$ the above model of opinion dynamics takes inspiration from the model of Weissbuch-Deffuant \cite{Weissbuch:2000} or related kinetic exchange models \cite{Lallouache:2010,Sen:2011}, but, similar to the DeGroot model \cite{DeGroot:1974}, treats the dynamics of opinion change as synchronous updates accounting for simultaneous influences of all neighbours instead of subsequent pairwise interactions as in \cite{Weissbuch:2000}. 

In the following, we will consider a simultaneous dynamics of opinion change and network change in which individual agents competitively attempt to rewire connections to enhance their influence on the overall consensus state. For this purpose, we will assume that each agent can change who she/he influences (i.e. adjust her/his out-connections), but note that we do not allow for self-connections. The influence of an agent on the system can be measured by the closeness of the average opinion $X=1/N\sum_j x_j$ to the agents opinion, i.e. the influence of agent $i$ is given by
\begin{equation}
\label{E1}
    I_i = \left| x_i- X \right|,
\end{equation}
where $i$ may be considered as having maximum influence if $I_i=0$, i.e. the system completely agrees with $i$ and minimum influence $I_i \approx 2$ if all other agents maximally disagree from $i$.

In more detail, we consider the following adaptive dynamics \cite{Blasius:2008} of opinion change and network evolution: (i) opinions evolve for one time step according to Eq. (\ref{E0}). Boundary conditions for the opinion space are strictly enforced, such that opinions will always be restricted to the interval $[-1,1]$. Note, that we have also experimented with asynchronous updating in which, instead of simultaneously updating the opinions of all agents, only a number of randomly selected agents update their state according to Eq. (\ref{E0}). The main effect of such asynchronous updating is a change of time scales of opinion formation relative to network change. Beyond rescaling of time, a change to asynchronous updating does not result in qualitative differences to the results we present below for synchronous updating. (ii) Subsequently, $m$ agents are chosen at random. Each of them considers rewiring an out-link $x\to y$ to $x \to z$, where $z$ is selected at random from all nodes with $z \not=x$ and $a_{zx}=0$. Specifically, agent $x$ will rewire, if her/his influence $I_x(\Gamma)>I_x(\Gamma^\prime)$ is improved in the rewired network configuration $\Gamma^\prime$ compared to her/his influence in the original network configuration $\Gamma$. To estimate her/his influence in both configurations, agent $i$ will estimate opinions in the next time step based on Eq. (\ref{E0}) ignoring unpredictable external effects (i.e. setting $\Delta=0$). If the rewiring is accepted, we proceed with the rewired configuration $\Gamma^\prime$, otherwise the network is restored to its original state $\Gamma$. (iii) The iteration of steps (i) and (ii) is then repeated until a quasi-stationary state has been reached.

Note that both the parameters $m$ and $\alpha$ essentially define the time scale of network change relative to the consensus dynamics, which is slow for small $m$ (or large $\alpha$) and increases in relative speed when $m$ increases (or $\alpha$ decreases). For numerical reasons we will fix the value of $\alpha$ (using a reasonably slow time scale of opinion dynamics with $\alpha=10^{-4}$ such that network adaptation can have an effect before a consensus is reached) and mainly explore the dependence of outcomes on changes of $m$ below.  

The above dynamics essentially describes a competitive process in which agents subsequently attempt to change the system trying to align it with their own opinions. As agents typically hold different opinions, their goals are typically not aligned. Note, that such a scenario could also be formalized as a network creation game with $N$ actors and one could attempt to find game-theoretic equilibria of this game. The resulting strategy space in such a game would be of enormous size and strict computation of equilibria would require extremely large computational resources on the part of the actors. Here instead, we assume boundedly rational agents with limited computational power, and consider the above defined best-response-like dynamics, similar to \cite{Brede:2010,Brede:2018c}.

Below, we are mostly interested in consensus formation in the population of agents. In the proposed model consensus is well characterized by the spread of opinions and hence we define an order parameter 
\begin{equation}
    r^2=1/N \sum_i (x_i-\langle x\rangle)^2
\end{equation}
such that $r=0$ corresponds to a perfect consensus. An important measure is also the time scale of equilibration towards consensus. Following \cite{Hong:2002,Son:2008,Brede:2010}, we measure this scale by defining a normalized order parameter
\begin{equation}
 \tilde{r(t)}=\frac{r(t)-r_\infty}{r_0-r_\infty}
\end{equation}
and then measure a relaxation time scale via the sum
\begin{equation}
 \tau_\text{relax} = \sum_{t^\prime=0}^\infty \tilde{r} (t^\prime), 
\end{equation}
which can capture the effect of multiple relaxation times for a monotonically decaying function.

\section{Results}
\label{results}
In the following numerical simulations are carried out to investigate the effects of the adaptive dynamics of opinion spread and competitive influence maximization on consensus formation. In principle, we are interested in two types of settings. In the first, we will consider an endogenous adaptive dynamics in which a large fraction (or all) of the population of agents simultaneously attempt to maximize their influence on the system. In this setting, all agents are considered endogenous, and they can thus be influenced by others as well as influence others. We will analyze this setting in subsection \ref{Results:Endo}. The second setting of interest is the exogenous setting, in which influencers are extrinsic to the system. In this case, these external agents can influence (and optimize their influence on) the system, but they are not themselves subject to influence. We will analyze this scenario in subsection \ref{Results:Exo}.

\subsection{Competitive rewiring: Endogenous dynamics}
\label{Results:Endo}
\begin{figure}[tbp]
\begin{center}
\includegraphics[width=.45\textwidth]{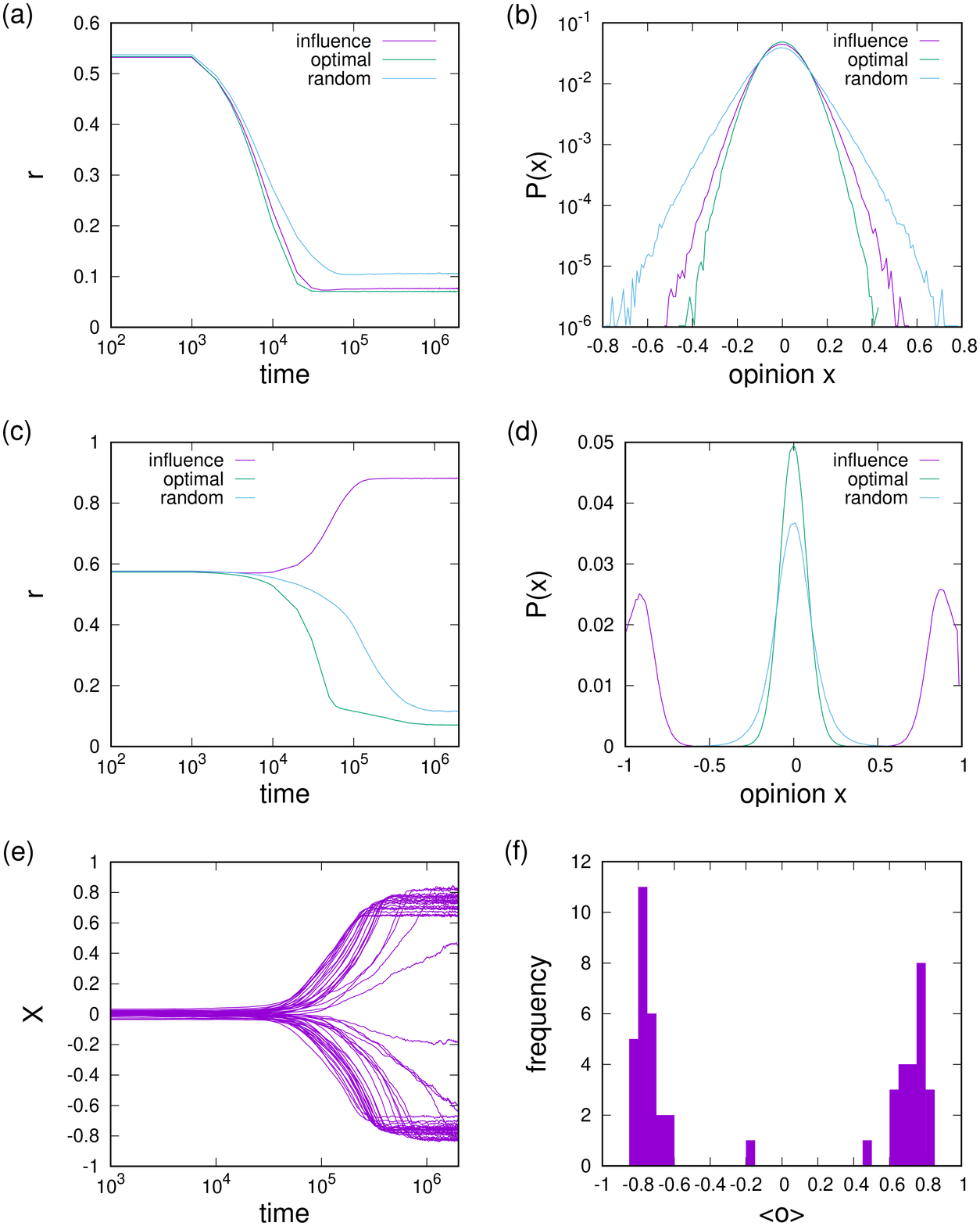}
\caption {Different equilibrium outcomes for the influenced opinion dynamics for different settings of the tolerance parameter $\delta$: (a,c) averaged time dependence of the opinion dispersion $r$ for $\delta=1.5$ and $\delta=0.5$, (b,d): stationary distributions of the opinions in equilibrium for $\delta=1.5$ and $\delta=0.5$. The figures compare three different settings for rewiring: competitive influence maximization (influence), rewiring to reduce $r$ (optimal), and random rewiring (random). (e,f): $\delta=0.75$. (e) Evolution of the average opinion $X$ for various example runs and (f) distribution of the final averaged opinion $X$. Comparisons to non-competitive dynamics omitted in (e) and (f). Results are averaged over $50$ independent runs for a system of $N=1000$ nodes, noise $\Delta=0.001$, $\langle k\rangle=10$, and $m=1$. }
\label{fig.0}
\end{center}
\end{figure}

\begin{figure*}[tbp]
\begin{center}
\includegraphics[width=.95\textwidth]{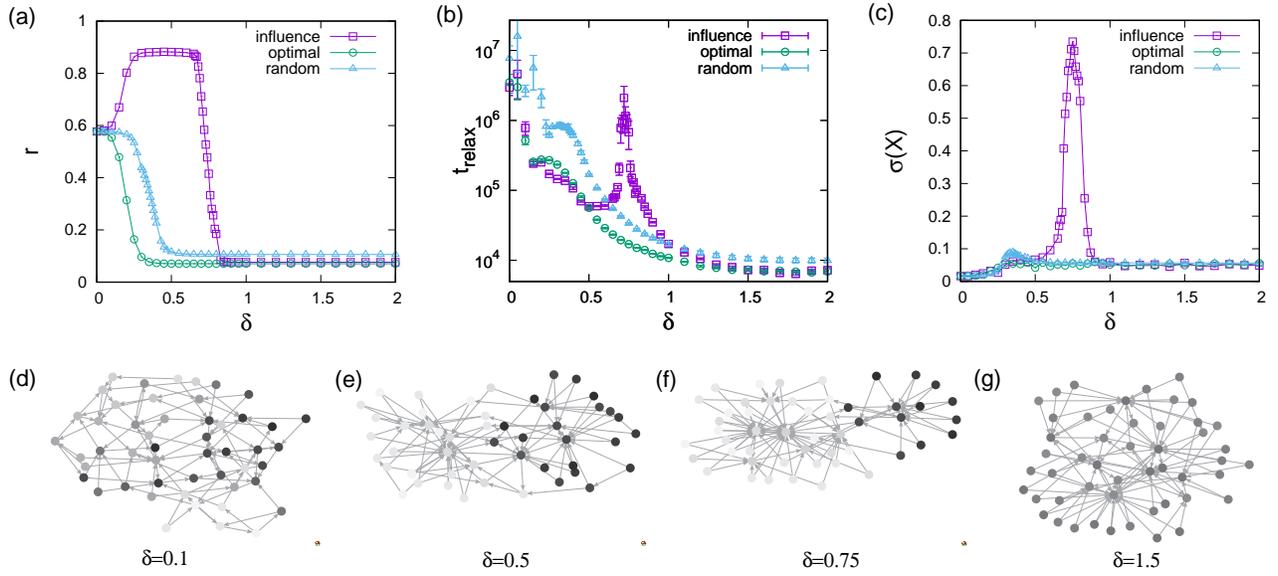}
\caption {Dependence of outcomes of the adaptive influence dynamics on $\delta$ for a scenario of all agents attempting to maximize their influence on the system: (a) opinion dispersion, (b) relaxation times, (c) standard deviation $\sigma(X)$ of the average opinion X at the end of simulations. The figures compare three different settings for rewiring: competitive influence maximization (influence), rewiring to reduce $r$ (optimal), and random rewiring (random). Results are averaged over $50$ independent runs for a system of $N=1000$ nodes, noise $\Delta=0.001$, $\langle k\rangle=10$, $m=3$. The bottom row visualizes typical evolved networks of size $N=50$ with $\langle k\rangle=4$ for $\delta=0.1$ (left most), $\delta=0.5$ (2nd from left), $\delta=0.75$ (third from left) and $\delta=1.5$ (right). Nodes are coloured according to their opinions with light gray corresponding to $x=-1$, gray to $x=0$, and black to $x=1$.}
\label{fig.1}
\end{center}
\end{figure*}

\begin{figure}[tbp]
\begin{center}
\includegraphics[width=.45\textwidth]{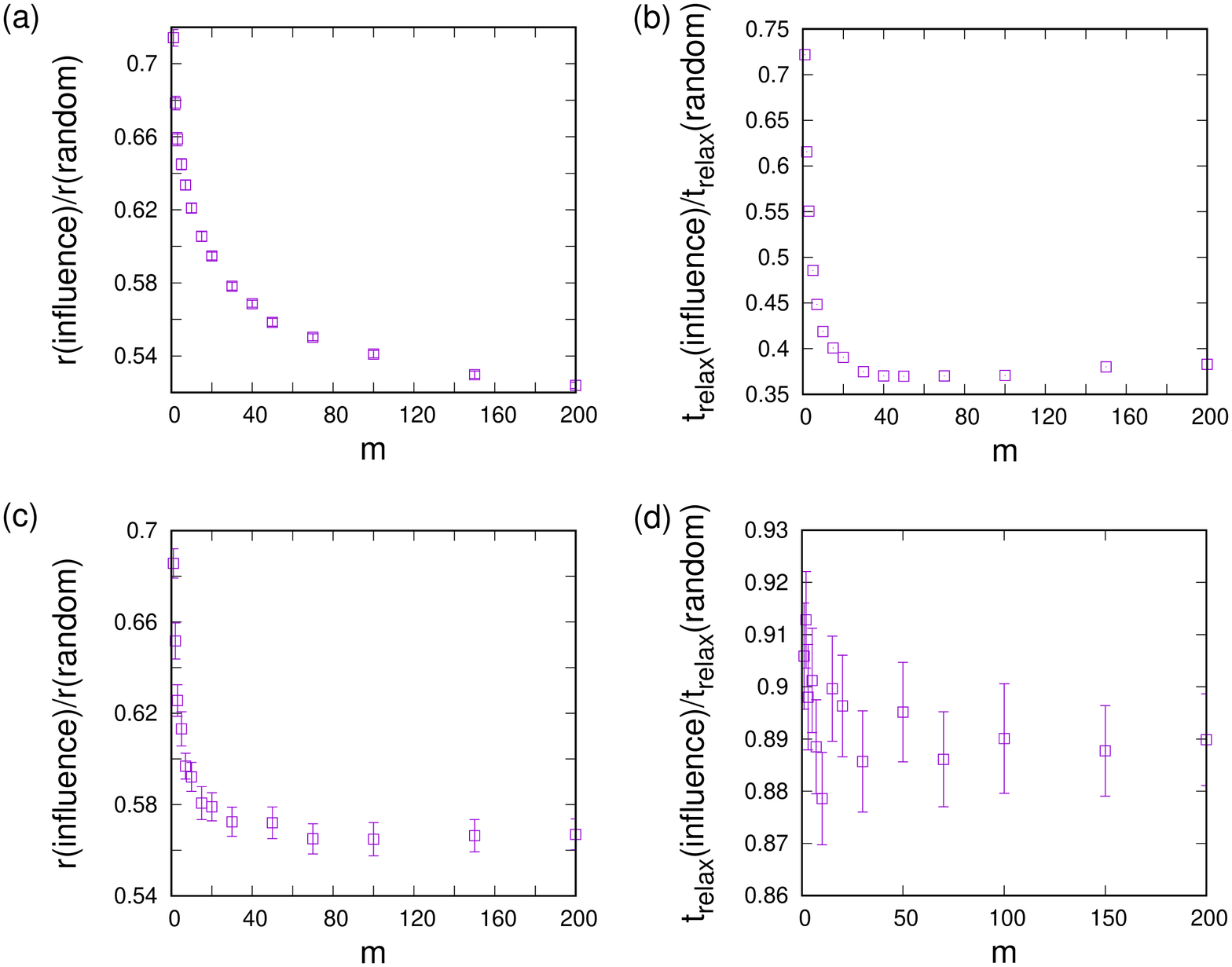}
\caption {Dependence of (a,c) the average relative opinion dispersion $r$ and (b,d) average relative relaxation times on the intensity of influence maximization (measured by the number of attempted optimization steps $m$ per update) for the scenario of a network of all agents attempting to maximize their influence on the system (top) and a scenario of two external agents with opinions $x=\pm 1$ attempting to maximize their influence on the system. Results are averaged over $50$ independent runs for a system of $N=1000$ nodes, noise $\Delta=0.001$, $\langle k\rangle=10$. In the scenario of external influencers each controls $k_\textrm{ext}=509$ nodes.}
\label{fig.2}
\end{center}
\end{figure}

Figure \ref{fig.0} provides some first numerical results illustrating the adaptive dynamics of opinion change and network rewiring for three settings of low ($\delta=0.5$), intermediate ($\delta=0.75$) and high ($\delta=1.5$) tolerance parameter $\delta$. For comparison, also results obtained for two different adaptive rewiring processes are included. In the first we consider purely random rewiring, i.e. in step (ii) of the adaptive dynamics every rewired configuration is accepted independent of whether it enhances influence or not. Random rewiring is of interest to evaluate the role of network change and serves as a baseline for comparison. As a second process, we have also considered rewiring in which configurations are accepted if they reduce the overall spread of opinions $r$. We label this process "optimal", as it represents a rewiring dynamics that aims for the fastest convergence of opinions to a state with the lowest possible spread.

Panel (a) of Fig. \ref{fig.0} illustrates the time evolution of opinion spread $r$ for large $\delta$ and compares the three adaptive rewiring processes. We see that subject to all three adaptive dynamics an initially diverse distribution of opinions collapses over time and finally reaches a quasi-stationary state after an initial transient period. Results in Fig. \ref{fig.0}a show that the adaptive dynamics with random rewiring is relatively slow and leads to a more spread out distribution of opinions than optimal and influence maximizing rewiring, cf. also panel (b) in which the stationary distributions of opinions are compared. Importantly, we see that for influence maximizing rewiring the consensus dynamics proceeds at almost the same time scale as that of the optimal process, resulting in an only slightly more spread out distribution than when deliberately minimizing spread. The figure already highlights a key result of our study, i.e. that competitive influence maximization can speed up consensus formation and  can result in near-optimal stationary consensus states.

However, panels (c) and (d) of Fig. \ref{fig.0} illustrate that influence maximization does not always accelerate consensus formation. In the figure we see that for small tolerance parameters competitive influence maximization can also result in opinion `fragmentation', i.e. states in which the population is split into two `radicalized' clusters of agents holding opinions close to the boundaries of the opinion space. Note, that similar clustering is not observed in the cases of random rewiring or the optimal rewiring process. In both cases, for the given tolerance parameter $\delta=0.5$, a unimodal distribution of opinions centred around the middle of the interval $[-1,1]$ is reached (cf. Fig. \ref{fig.0}d).

The case of intermediate tolerance $\delta$ is illustrated in panels (e) and (f) of Fig. \ref{fig.0}. In panel (a), we show an ensemble of trajectories for the evolution of the average opinion $X=1/N\sum_i x_i$ and also plot the corresponding histogram over final average opinions at the end of the simulations. We note clusters of $X$ at around $X=\pm 0.8$ implying that also for this value of $\delta$ opinions become polarized with a majority of agents either approaching the left or right boundary of the opinion space, whilst a minority converges to the respective opposite end of the opinion space. The situation reflects a `winner-takes-most' scenario, in which, with equal chance, either radical end of the opinion space can grow to dominance. The extent, to which this is the case is quantified in the distribution of final average opinions $X$: the larger and the farther away from each other the peaks of this bimodal distribution, the stronger the tendency for the winner-takes-most scenario. We can thus quantify the severity of the winner-takes-most effect via the standard deviation $\sigma(X)$ of the distribution of average opinions. We find that small standard deviations correspond to a narrow distribution of outcomes centred around a consensus in the middle of the interval $[-1,1]$ and large standard deviations correspond to the bimodal distribution with peaks close to $\pm 1$. As we will see below, winner-takes-most scenarios are not encountered for random or optimal rewiring.

In Figure \ref{fig.1} we systematically explore the dependence of the adaptive dynamics on the tolerance parameter $\delta$. We show simulation data for averaged opinion spreads $r$ (panel a), averaged relaxation times towards the stationary state (panel b) and the standard deviation of average opinions (panel c), and again compare the influence maximization to random rewiring and optimal rewiring. The bottom row of the figure visualizes typical network configurations for very small, small, intermediate, and large choices of $\delta$. 

First, for very small tolerance parameters, we find configurations marked by the presence of several opinion clusters, cf. also the network visualization in panel (d). Convergence times in this regime are typically very long, see Fig. \ref{fig.1}b. Second, as $\delta$ is increased, configurations reached subject to influence maximization exhibit marked differences to random or optimal rewired networks. In this regime, we typically find opinion polarization and radicalization with two clusters close to the borders of the opinion space arising, see also panel (e) for a visualization of such a network. It is worth pointing out that such radical polarization is not found in either random or optimal rewiring and is thus caused by competitive rewiring. 

Further increasing $\delta$ the regime of polarization and radicalization ends in a third regime, a fairly small transition region, in which typically still two radicalized clusters at opposite ends of the opinion space appear. However, opinion spreads are lower than in the previous regime, corresponding to the two clusters having markedly different sizes. In this regime we find the winner-takes-most effect described above. This transition region is well demarcated by a corresponding peak in the plot of the standard deviations of mean opinions in panel Fig. \ref{fig.1}c as well as by a peak in convergence times in Fig. \ref{fig.1}b. For yet larger $\delta>1$ a fourth regime, in which compromise consensus states marked by very low opinion spreads $r$ are reached, is found. In this regime we note the near-optimal opinion spread $r$ and close-to-optimal convergence times, as discussed above. For a more detailed discussion of issues of scaling with system size or different choices of noise we refer to subsections Ia and Ib of the supplementary material. 

\begin{figure}[tbp]
\begin{center}
\includegraphics[width=.45\textwidth]{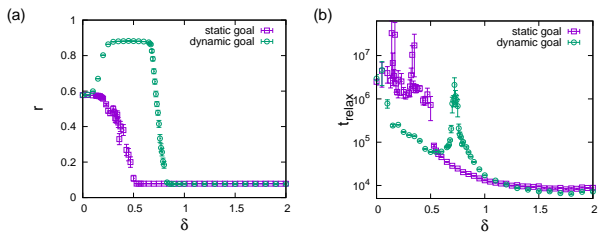}
\caption {Comparison between agents which attempt to maximize the influence of their (dynamic) opinions with agents which optimize the influence with regard to static randomly drawn goals (static). Dependence of (a) the averaged opinion spread and (b) average relaxation times on the tolerance parameter $\delta$. Results are averaged over $50$ independent runs for a system of $N=1000$ nodes, noise $\Delta=0.001$, $\langle k\rangle=10$, $m=3$.}
\label{fig.22}
\end{center}
\end{figure}

Interestingly, consensus enhancement in the large $\delta$ region is not dependent on the feedback of agents maximizing the influence of their own (dynamically changing) opinions. To test the impact of this feedback, we also considered a model in which agents attempt to maximize influence defined by
\begin{equation}
\label{E1a}
    I_i = \left| g_i- X\right|,
\end{equation}
where $g_i$  is a randomly drawn goal state $g_i \in [-1,1]$ instead of an agents' own actual opinion as in Eq. (\ref{E1}). Figure \ref{fig.22} compares simulations based on the two different influence definitions following equations (\ref{E1}) and (\ref{E1a}). Two observations stand out. First, in the second scenario, we note the absence of the polarized radicalized state for low $\delta$ and the absence of the winner-takes-most states. Second, for large $\delta$ in the second scenario competitive influence maximization also reaches a slightly worse, but still near-optimal consensus state with slightly slower transients compared to the scenario in which agents maximize the influence of their actual opinions. We can thus conclude that the radicalizing polarisation and the winner-takes-most effect are caused by the feedback of adapting opinions on influence maximization. In contrast, consensus enhancement and accelerated transients can be achieved when agents attempt to maximize influence, irrespective of the actual goals they pursue.

As we have seen in Fig. \ref{fig.1}, depending on the tolerance parameter, competitive influence maximization can either result in radicalization and polarization or it can facilitate consensus formation and result in a far lower spread consensus states than random rewiring. The exact spread of the stationary consensus state depends on the relative time scales of adaptive influence maximization and the opinion dynamics. More specifically, the more agents $m$ are allowed to rewire per time step of the opinion dynamics, the more important competitive influence maximization becomes relative to the dynamics of opinion spread. For large $\delta$ we explore this dependency in more detail in the panels of Fig. \ref{fig.2} where we visualize results obtained from experiments in which $m$ was varied between $m=1$ rewiring agent per time up to $m=200$ rewiring agents per opinion update. As more intense random rewiring tends to improve mixing and will thus also result in narrower consensus states, to factor out the latter effect, results in Fig. \ref{fig.2} are given relative to random rewiring with the same number $m$ of rewiring agents per opinion update. We can clearly see that more intense influence maximization will result in relatively narrower consensus states (Fig. \ref{fig.2}a)  and typically also in shorter transients and faster dynamics to reach the consensus state (see Fig. \ref{fig.2}b).  We have also conducted analogous experiments for the low $\delta$ setting for which we don't illustrate results. As one would expect, in the this setting we also find that equilibration to the polarized configurations becomes the faster the larger $m$.

A next question that arises is how results of the competitive influence maximization process depend on the number of participating agents. Do consensus states become the more sharply focused the more agents attempt to impose their opinions? To explore this issue we conduct experiments in which we randomly select a proportion $\rho$ of agents as active. Whereas all agents continue to be subject to opinion changes specified in Eq. (\ref{E0}) only active agents will attempt to rewire their out-connections in the influence-maximizing step of the adaptive dynamics, i.e. when choosing an agent for rewiring we restrict the choice to randomly selecting an active agent.

\begin{figure}[tbp]
\begin{center}
\includegraphics[width=.45\textwidth]{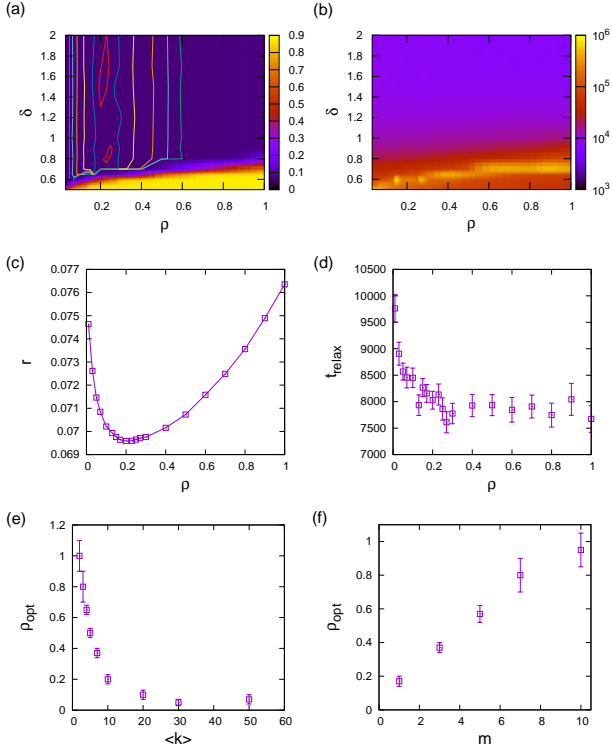}
\caption {Dependence of (a) the stationary spread $r$ and (b) the convergence times $t_\textrm{relax}$ on the tolerance parameter $\delta$ and the density of influence optimizing agents $\rho$. For (a) isolines at $r=0.0696, 0.0697, 0.0700, 0.0705, 0.0710, 0.0715$ have been added to visualize the optimum at $\rho \approx 0.2$. (c) and (d) show the dependence of $r$ and $t_\textrm{relax}$ on $\rho$ for $\delta=1.5$. Panels (e,f) show the dependence of the optimal proportion of influence-maximizing agents on (e) the average connectivity of the social network $\langle k\rangle$ and on (f) the intensity of influence maximization $m$. Results are averaged over $50$ independent runs for a system of $N=1000$ nodes, noise $\Delta=0.001$, $\langle k\rangle=10$, $m=1$. }
\label{fig.da}
\end{center}
\end{figure}

Results of numerical simulations that explore the dependence of stationary outcomes on $\rho$ and the tolerance parameter $\delta$ are shown in Fig. \ref{fig.da}. Panel (a) of Fig. \ref{fig.da} visualizes a heat map of stationary opinion spreads as a function of $\rho$ and $\delta$ and panel (b) gives the corresponding heat map for relaxation times. We note the existence of the two regimes discussed before: i.e. a regime of two polarized radical clusters for low $\delta$ and a compromise consensus state for large $\delta$, separated by a transition region in which relaxation times become very large. Two observations are in order. First, the location of the boundary between the regimes depends on $\rho$. The smaller the proportion of influence maximizing agents, the lower the tolerance threshold at which polarization sets in (see Fig. \ref{fig.da}a). Second, as illustrated by the isolines in the large compromise consensus regime, a reduction in rewiring agents does not necessarily impede consensus formation. Panel (c) of Fig. \ref{fig.da} makes it clear that there exists an optimal number of influence maximizers at around $\rho=0.2$ which minimizes opinion spread. As we see in panel Fig. \ref{fig.da}d, such minimal opinion spread can be reached without much loss in convergence time, i.e. convergence times have typically plateaued off in the $t_\textrm{relax}(\rho)$-dependence before the optimal value of $\rho$ has been reached. We thus find that the best compromises are reached when only a certain proportion of agents actively tries to maximize their influence on the system.

It is of interest to investigate which factors influence the optimal number of rewiring agents $\rho_\textrm{opt}$. For instance, from the isolines in Fig. \ref{fig.da}a which are parallel to the y-axis we see that $\rho_\textrm{opt}$ is independent of the tolerance parameter $\delta$. We have carried out extensive numerical simulations to explore $r(\rho)$ dependencies for different settings of noise strength $\Delta$, social network connectivities, and different intensities of influence maximization $m$. Whereas the first parameter has no influence on $\rho_\textrm{opt}$, we find a clear dependence on the connectivity of the social network and the intensity of influence maximization $m$. These dependencies are illustrated in panels (e) and (f) of Fig. \ref{fig.da}. Panel (e) illustrates simulation data for different connectivities of the social networks. We see that the more connected the social network, the smaller the optimal fraction of maximizing agents. That is, optimal consensus states on sparse social networks can be achieved when large proportions of the population attempt to maximize influence, whereas optimal results in dense networks require only very limited participation in influence maximization. The relationship between optimal participation $\rho_\textrm{opt}$ and intensity of influence maximization is explored in panel (f) of Fig. \ref{fig.da}. We note that the more intense the influence maximization relative to the dynamics of opinion spread, the larger the optimal participation ratio. 

\begin{figure}[tbp]
\begin{center}
\includegraphics[width=.45\textwidth]{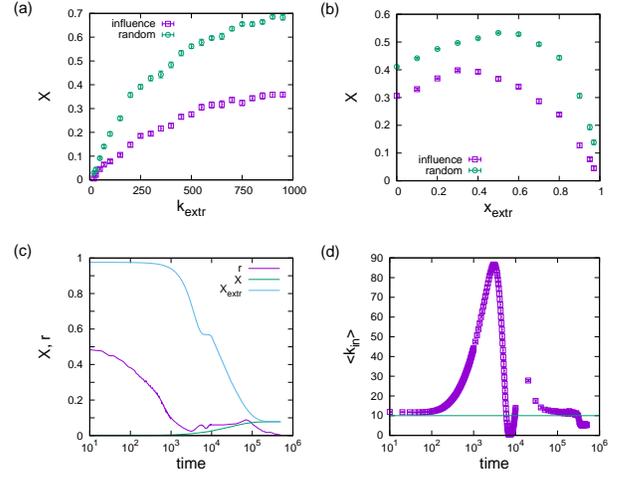}
\caption {Dependence of mean consensus $X$ on the presence of influential minorities. (a) Dependence of $X$ on the out-degree of minority agents $k_\textrm{extr}$ in a setting where normal agents have out-degree $k_\textrm{out}=10$. Minority agents are defined as agents with initial opinions $x_i(0)>0.95$. (b) Dependence of mean consensus $X$ on a threshold $x_\textrm{extr}$ in a setting in which normal agents have $k_\textrm{out}=10$ and agents with initial opinion $x_i(0)>x_\textrm{extr}$ are assigned out-degree $k_\textrm{out}=100$.  (c) Evolution of the spread of the mean $X$ for all agents and mean $X_\textrm{extr}$ for the minority group and spread of consensus state $r$ over time. (d) Evolution of the in-degree $k_\textrm{in}$ of minority agents over time.  Results are averaged over $20$ independent runs for a system of $N=1000$ nodes, noise $\Delta=0$, $\langle k\rangle=10$. To emphasize the effects of influence maximizing rewiring, the time constant for opinion convergence is chosen one order of magnitude smaller than in the rest of the paper ($\alpha=10^{-5}$) and we use a large value of $m=100$. Panels (c) and (d) are for $x_\textrm{extr}=0.95$ and $k_\textrm{extr}=100$.}
\label{fig.6}
\end{center}
\end{figure}

A last point of interest in this section is to investigate the role of influential minorities on consensus formation in the presence of competitive influence maximization. For this purpose, we consider two types of numerical experiments. In the first, we introduce an influential minority by allocating larger out-degrees of $k_\textrm{out}>10$ to agents with initial opinions $x_i(0)>0.95$, whereas all other agents are allocated out-degrees of $k_\textrm{out}=10$. We thus set up an influential minority of agents with extreme opinions which comprises only $2.5\%$ of the agent population. We then vary $k_\textrm{extr}$ in the interval $[10,1000]$, allowing the minority to control between $2.5\%$ and roughly $72\%$ of all links. It is thus expected that influential agents with $k_\textrm{out}>10$ would have a larger influence on consensus states than normal agents with $k_\textrm{out}=10$ and we expect average consensus states to be significantly shifted into the positive domain. 

Simulation results for this first type of experiment are visualized in Fig. \ref{fig.6}a where we compare the dependence of average consensus states for influence maximizing rewiring and random rewiring. Note that, in order to rule out random drift of the extreme minority, we have not included noise in opinion updates for these experiments. In Fig. \ref{fig.6}a we observe that in both cases $X>0$ and we observe a monotonic increase in $X$ with increasing influence $k_\textrm{extr}$ of minority agents, as expected. However, it stands out that consensus states are significantly less strongly impacted by the minority opinion in the presence of influence maximizing rewiring. 

In a second type of setting, we have allocated minority agents larger out-degrees $k_\textrm{out}=100$, but defined minority agents as agents with initial opinions $x_i(0)>x_\textrm{extr}$ and then varied $x_\textrm{extr}$, allowing to tune the fraction of minority agents in the population. As initial opinions are drawn at random from $[-1,1]$ we have varied $x_\textrm{extr}$ between $x_\textrm{extr}=0$ and $x_\textrm{extr}=1$, changing their share in the overall population from close to 0\% to 50\%. Numerical results for this setting are given in Fig. \ref{fig.6}b, where we again compare to adaptive random rewiring. We note again that the presence of an extreme minority sways consensus states, but the influence of the minority is attenuated when adaptive influence maximizing rewiring is present.

To understand the mitigating influence of competitive influence maximization we have tracked the evolution of average opinions of the population and of the minority and the in-degrees of the influential minority agents in Figures \ref{fig.6}c,d. Figure \ref{fig.6}d makes it clear that before getting close to reaching a consensus state within the population minority agents attract substantially above average in-degrees. More detailed observation shows a wave like pattern in which the minority initially attracts large in-degrees, but loses and gains in-links as it holds or loses the position of representing the most extreme opinions in the population. These results highlight two feedback mechanism inherent in the process of competitive influence maximization. First, within the tolerance range $\delta$, agents with opinions at the extremes of the distribution of opinions in the population attract more in-links and thus tend to be drawn faster toward the consensus state. This mechanism is the crucial feedback allowing populations to reach sharper consensus states with competitive influence maximizing rewiring. Second, since by influencing influential agents other agents can enhance their respective influence, also agents which are more influential than average tend to attract larger than average numbers of in-links. Influential extreme minorities are subject to both aspects of this feedback mechanism and their opinions tend to be quickly drawn towards the average of the population, such that the number of neighbours they can convert to extreme views remains limited.

\subsection{Competitive rewiring: The exogenous setting}
\label{Results:Exo}

\begin{figure}[tbp]
\begin{center}
\includegraphics[width=.45\textwidth]{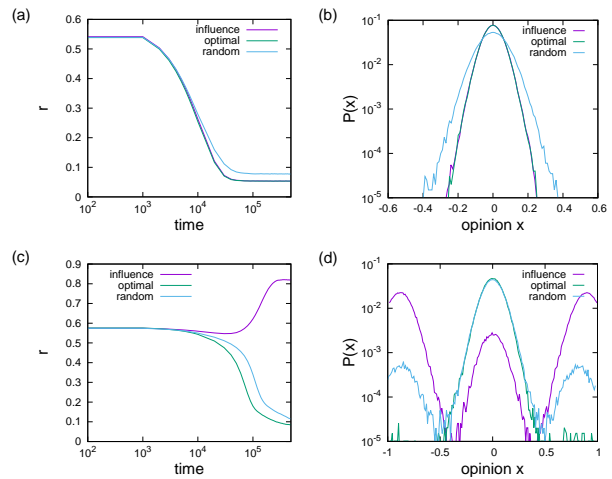}
\caption {Different equilibrium outcomes for the influenced opinion dynamics for different settings of the tolerance parameter $\delta$ for the scenario of two external controllers with opinions $\pm 1$: averaged time dependence of the opinion dispersion $r$ for (a) $\delta=1.5$ and (c) $\delta=0.6$ and stationary distributions of the opinions in equilibrium for (b) $\delta=1.5$ and (d) $\delta=0.6$. The figures compare three different settings for rewiring: competitive influence maximization (influence), rewiring to reduce $r$ (optimal), and random rewiring (random). Results are averaged over $50$ independent runs for a system of $N=1000$ nodes, noise $\Delta=0.001$, $\langle k\rangle=10$, $m=1$. }
\label{fig.0a}
\end{center}
\end{figure}

\begin{figure*}[tbp]
\begin{center}
\includegraphics[width=.95\textwidth]{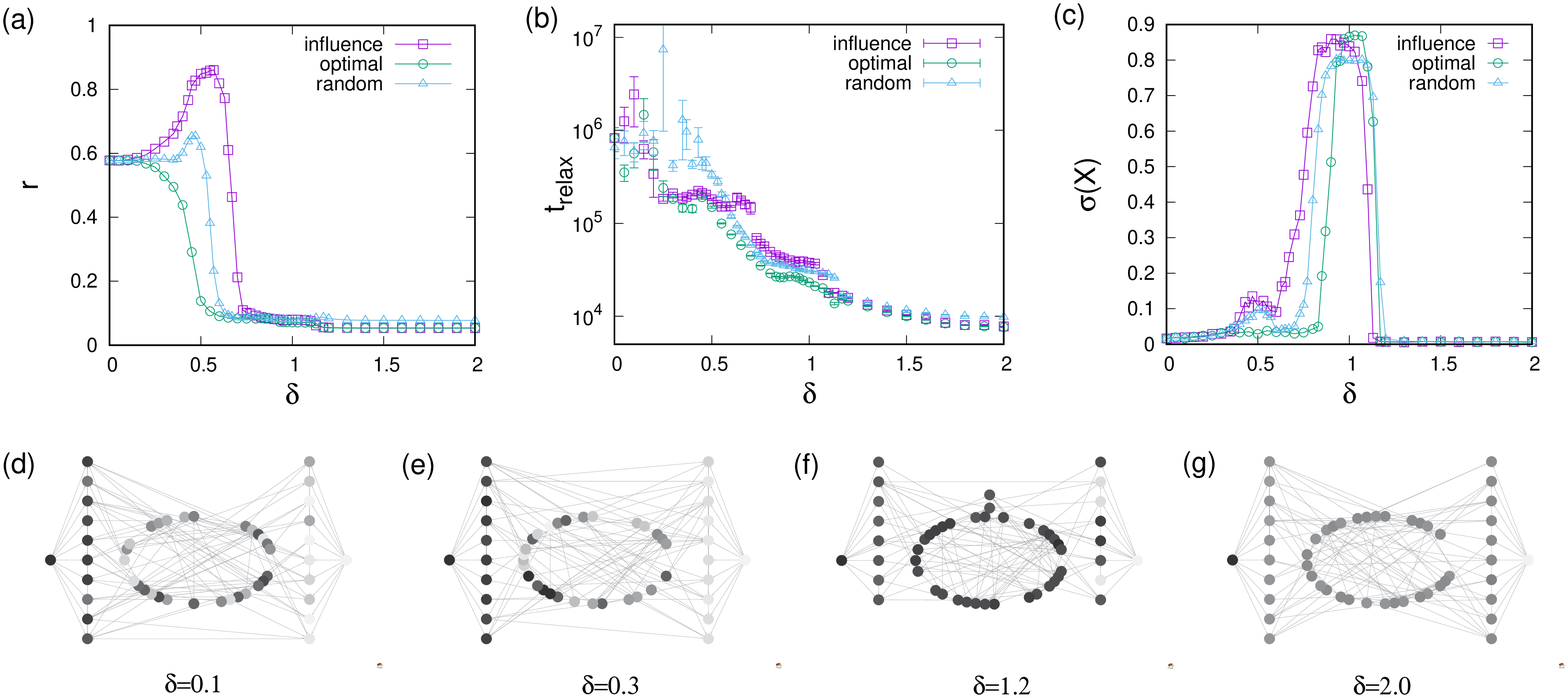}
\caption {Dependence of outcomes of the adaptive influence dynamics on $\delta$ for a scenario of two external controllers with opinions $\pm 1$ attempting to maximize their influence on the system: (a) opinion dispersion, (b) relaxation times, (c) standard deviation $\sigma(X)$ of the average opinion X. The figures compare three different settings for rewiring: competitive influence maximization (influence), rewiring to reduce $r$ (optimal), and random rewiring (random). Results are averaged over $50$ independent runs for a system of $N=998$ nodes noise $\Delta=0.001$, $\langle k\rangle=9$. The external controllers have out-degrees $k_\textrm{ext}=509$, so that the overall network has average degree $\langle k\rangle=10$ and we use $m=1$. The bottom row shows example network configurations for a network composed of 50 nodes with $\langle k\rangle=4$ and external influencers controlling $10$ nodes each. External controllers are indicated as black or white agents at the boundaries of each network visualization. Nodes directly controlled by the two external controllers are aligned along vertical lines next to the controllers, whereas nodes that are not directly controlled are positioned along an ellipse in the centre of each network visualization. The figures are for settings of $\delta=0.1$ or $\delta=0.3$  (first and 2nd from left, two opinion clusters), $\delta=1.2$ (middle, one opinion has won) and $\delta=2.0$ (right, consensus has been reached as a compromise). }
\label{fig.1a}
\end{center}
\end{figure*}

In this section we consider scenarios only involving external influencers. Note, that in contrast to Sec. \ref{Results:Endo} the external controllers are not subject to influence from other agents.

We start the analysis by considering two external influencers which hold opinions $x=-1$ and $x=1$ at the borders of the opinion space. Network parameters are adjusted such that $\langle k\rangle=9$ and the external influencers both have out-degree $k_\textrm{ext}=509$, such that the effective connectivity $\langle k\rangle=10$ is the same as in the endogenous setting.

Figure \ref{fig.0a} compares the competitive rewiring dynamics for this setting to a scenario in which the external influencers either rewire randomly or with an aim of minimizing opinion spread. Similar to what we have seen for the case of endogenous competitive influence maximization, we again note that for large tolerance $\delta$ competitive influence maximizing rewiring of the now external controllers can lead to a near-optimal consensus state which can be reached in near-optimal time, cf. panels \ref{fig.0a}a and \ref{fig.0a}b. In contrast, for low tolerance, a three-cluster state with one compromise-cluster centred around $x=0$ and two clusters aligned with the opposite controllers at $x=\pm 1$ (cf. panel (d) of Fig. \ref{fig.0a}), is found. Hence, interestingly, even though the two controllers hold extreme opposite opinions, their respective attempts to gain influence can actually promote a compromise consensus within the population, given that the tolerance parameter $\delta$ is large enough. To understand this effect, consider the case of $\delta=2$ and a population of agents in the absence of a social network. Since, according to the opinion dynamics in Eq. (\ref{E0}), $\Delta x_i (t) \propto x_j(t)-x_i(t)$, a controller $j$ can then maximize its influence by targeting nodes $i$ which maximize $\mid x_j(t)-x_i(t) \mid$. The controller with $x=1$ will thus target agents as close to $x=-1$ as possible, and the controller with $x=-1$ will target agents as close to $x=1$ as possible. In both cases radical agents from the opposite end of the spectrum are targeted and their opinions are drawn closer toward the centre of the opinion space. Thus, as a result of the competitive influence maximization of the two external controllers, the most radical agents are selected in turn and moved closer to the centre, thus resulting in improved consensus states. In contrast, for small $\delta$ when the ranges of the two external controllers don't overlap, agents at the borders of their respective ranges will be targeted, resulting in two radicalized clusters at the boundaries of the opinion space while agents in the centre remain uninfluenced, thus explaining the presence of the three cluster states noted in numerical results above.

\begin{figure}[tbp]
\begin{center}
\includegraphics[width=.45\textwidth]{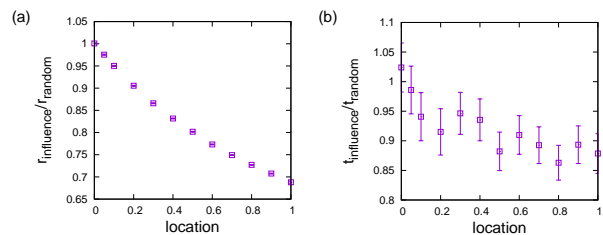}
\caption {Dependence of the relative opinion spread $r_\textrm{influenc}/r_\textrm{random}$ and relative relaxation times $t_\textrm{influence}/t_\textrm{random}$ on the distance of the location of the external influencers from $x=0$. Locations at $x=0$ correspond to two external influencers holding identical opinions at $0$, whereas location $x=1$ corresponds to the two external influencers holding opinions $\pm 1$. Results are averaged over $50$ independent runs for a system of $N=1000$ nodes, $\delta=1.5$, noise $\Delta=0.001$, $\langle k\rangle=10$. }
\label{fig.4}
\end{center}
\end{figure}

We proceed with a numerical investigation of the case in which agents are connected by social networks. Figure \ref{fig.1a} shows the results of a systematic exploration of the dependence of outcomes of the adaptive dynamics on the tolerance parameter $\delta$ together with visualizations of typical network configurations. We note the following possible outcomes. For small $\delta$ the three-cluster state discussed above is found. As $\delta$ is increased, the opinion clusters at the boundaries of the opinion space grow, whereas the compromise cluster centred around $x=0$ shrinks. This regime corresponds to the rise in $r$ approaching a maximum at around $r\approx 0.85$ (cf. Fig. \ref{fig.1a}a). For the range $0.7 \leq \delta \leq 1.1$ we then find a regime in which one of the extreme clusters at $\mid x\mid =1$ grows to dominance and coexists with a smaller cluster at the opposite end of the opinion space. This corresponds to a very flat maximum in $\sigma(X)$ seen in Fig. \ref{fig.1a}c. It is noteworthy that the extent of this regime is far larger than for the endogenous setting (cf. Fig. \ref{fig.1}). Moreover, discrepancies in size of the two radicalized clusters in the endogenous setting tend to be far larger than in the endogenous setting, i.e. the larger cluster wins out almost completely such that a near consensus close to a radicalized state at one of the boundaries of the opinion space is reached. In contrast to the endogenous setting, also random rewiring and optimal adaptive rewiring support such a `winner-takes-most' regime. Beyond $\delta=1.1$ competitive rewiring results in a single compromise cluster around $x=0$ which can be reached in near-optimal transients.

As also observed in the endogenous regime, the spread of consensus states and time scales at which stationarity can be reached depend on the intensity $m$ of the competitive rewiring process relative to opinion spread. Results illustrated in the bottom panels of Fig.  \ref{fig.2} show that in the large tolerance regime the spread of consensus states decreases with the intensity of rewiring and also time scales decrease slightly with increasing $m$.

So far in this section we have analyzed the role of two external controllers located at the extreme boundaries of the opinion space. In the following, we parameterize the relative location of the external controllers by their distance $d$ from the centre of the opinion space at $x=0$ and explore the dependence of consensus states and time scales on $d$. Settings of $d=0$ thus correspond to the two controllers both attempting to align the system with an opinion of $x=0$ (i.e. both controllers pursue the same goal for this case), whereas a choice of $d=1$ reproduces the scenario of two controllers at $x=\pm 1$ investigated before. Intermediate values of $0\leq d\leq 1$ interpolate between both extremes. As also the reference case of random rewiring is influenced by the location of the controllers, we give results relative to random rewiring in the following. Numerical results for the relative stationary opinion spread and relative time scales to stationarity are visualized in the panels of Fig. \ref{fig.4}. We note that whereas relative time scales show only a slight tendency to decrease with $d$, consensus states become relatively sharper the farther apart the controllers are from each other. Thus, provided the distance of the controllers is small enough to remain within the high tolerance regime, consensus is facilitated the farther apart the external controllers.

\section{Summary and conclusions}
\label{summary}

In this paper we have investigated an adaptive network model of opinion formation that combines a dynamics of opinion change with competitive network rewiring of agents attempting to increase their respective influence on the rest of the population. The model is based on incremental adaptations of opinion states subject to network neighbours whose opinions differ less than a given bounded tolerance level $\delta$ and external effects modelled as white noise of strength $\Delta$. We have investigated the model in two settings, in which either all agents attempt to rewire and are subject to influence (the endogenous setting) or only external controllers not subject to influence from the system adaptively adjust their control (the exogenous setting).

We generally find that in both, endogenous and exogenous settings of competitive influence maximization, stationary outcomes of opinion polarization marked by the presence of extreme opinions at the boundaries of the opinion space and regimes of compromise, in which all agents adopt opinions close to an opinion representative of the initial average of the population, exist. In more detail, we have demonstrated the presence of three regimes. First, for low tolerance $\delta$, stationary outcomes are characterized by the presence of radicalized clusters at the extreme ends of the opinion space. Second, for intermediate tolerances, we find `winner-takes-most' states, in which one extreme opinion has grown to dominance, but coexists with the other extreme opinion in various proportions. Last, in a third outcome typical for large tolerances, agents end up in compromise states sharply focused around the centre of the opinion space. These results differ markedly from outcomes of homophily-driven adaptive rewiring in opinion dynamics, which exhibit different types of typically multi-cluster regimes and for which adaptive rewiring tends to render global consensus more difficult to achieve \cite{Kozma:2008}.

As one of our main results we have argued that in the third regime competitive influence maximization can suppress differences of opinions and result in very narrow distributions of stationary opinions which are close to results that could be achieved by rewiring deliberately aimed at generating consensus states. Moreover,  in the third regime, these states of near-optimally low opinion spread can also be reached within near optimally short transient times. Consensus states become the sharper and convergence time scales the shorter the more intense the competitive influence maximization process is relative to the process of opinion change. 

Interestingly, best consensus states are not achieved when all agents in the population attempt to maximize their influence. Instead, we show that there exists an optimal number of participating agents which will minimize opinion spread. We find that this number depends on both the intensity of the rewiring process and on the connectivity of the social network. Optimal spreads are achieved with smaller numbers of rewiring agents in dense social networks and rewire numbers of participating agents that are the larger the more intense the rewiring process relative to the dynamics of opinion change.

As a last important result, we have also shown that in the large tolerance regime competitive influence maximizing rewiring suppresses the influence of small influential minorities with extreme opinions. We have argued that under competitive influence maximization already more influential agents tend to attract large numbers of in-links. As a result such agents tend to be drawn back to average opinions and thus their impact on consensus states is reduced.

We think that our work also opens up a number of interesting direction for future work. First, we have limited our investigation to exploring processes of adaptive influence maximization separately in endogenous and exogenous settings. It will be of interest to also explore results in scenarios in which both processes are present at the same time and investigate the role of time scales at which internal and external agents rewire on the consensus dynamics. Second, our investigation in the present paper has been restricted to one specific model of consensus formation. Previous literature has investigated many variants of such models and some results may depend on details in the model formulation (c.f. \cite{Castellano:2009,Sirbu:2016} for reviews). For instance, we have treated opinions as continuous, thus allowing agents to gradually shift positions in response to influence. A large part of the literature has treated opinions as binary or discrete. Future work will also investigate such competitive influence maximization processes in discrete models of opinion formation.

\section{Acknowledgements}
The authors acknowledge the use of the IRIDIS High Performance Computing Facility, and associated support services at the University of Southampton, in the completion of this work. This work was supported by the University of Southampton and by the Turing-sponsored pilot project "Strategic influence in dynamic opinion formation".

\section{Conflicts of interest and data availability}
The author(s) declare(s) that there is no conflict of interest regarding the publication of this paper. Results described in this paper are based on numerical simulations described in the paper. No additional data were used to support this study.

\section{Supplementary Material}

In the supplementary material additional experiments regarding different choices of noise and the scaling with system size are discussed.
%\begin{figure*}[ht]
% \begin{center}
%\includegraphics[width=.45\textwidth]{nfigures/ts.eps}
% \caption {(a) .}
%\label{fig.0}
%\end{center}
% \end{figure*}

\bibliography{sample}
\end{document}